\documentclass[twocolumn,prl,amsmath,amssymb,aps,superscriptaddress]{revtex4-1}
\usepackage{graphicx}
\usepackage{dcolumn}
\usepackage{bm}
\usepackage{overpic}
\usepackage{xcolor}
\definecolor{MyBlue}{rgb}{0, 0.6353, 0.9098}
\definecolor{MyGreen}{rgb}{0.1333, 0.6941, 0.2980}
\newcommand{\ve}[1]{\ensuremath{\mbox{\boldmath$#1$}}}
\usepackage{amsmath,amssymb}

\newcommand{\be}{\begin{equation}}

\newcommand{\ee}{\end{equation}}
\begin{document}
\title{Finding Efficient Swimming Strategies in a Three-Dimensional Chaotic Flow by Reinforcement Learning}
\author{K. Gustavsson}
\affiliation{Department of Physics, University of Gothenburg,  Origov\"agen 6 B, 41296, G\"oteborg, Sweden.}
\author{Luca Biferale}
\affiliation{Department of Physics and INFN, University of Rome Tor Vergata, Via della Ricerca Scientifica 1, 00133, Rome, Italy.}
\author{Antonio Celani}
\affiliation{Quantitative Life Sciences, The Abdus Salam International Centre for Theoretical Physics, Strada Costiera 11, 34151, Trieste, Italy.}
\author{Simona Colabrese}
\affiliation{Department of Physics and INFN, University of Rome Tor Vergata, Via della Ricerca Scientifica 1, 00133, Rome, Italy.}
\email{simona.colabrese@roma2.infn.it}
\begin{abstract}
	\noindent
	We apply a reinforcement learning algorithm to show how smart particles can learn  approximately  optimal strategies to navigate in complex flows. In this paper  we consider microswimmers in a paradigmatic three-dimensional case given by a
	stationary superposition of two Arnold-Beltrami-Childress  flows with chaotic advection along streamlines. In such a flow, we study the evolution of point-like particles which can decide in which direction to swim, while keeping the velocity amplitude constant. We show that it is sufficient to endow the swimmers with
	a very restricted  set of actions (six fixed swimming directions in our case)  to have enough freedom to find efficient  strategies
	to move upward  and escape local fluid traps.
	The key ingredient is the learning-from-experience structure  of the algorithm, which assigns positive or negative rewards depending on whether the taken action is, or is not,  profitable for the predetermined goal  in the long term horizon.   This is another example supporting the efficiency of the reinforcement learning approach to learn how to accomplish difficult tasks in complex fluid environments.\footnote{\textbf{Postprint version of the article published on Eur. Phys. J. E (2017) 40: 110 DOI: 10.1140/epje/i2017-11602-9}}
\end{abstract} 
\maketitle
\section{Introduction}
\label{sec:introduction}
\noindent
Microswimmers can modify their motion in response to a physical or chemical stimulus from the surrounding environment.
They are equipped with simple sensory capacities 
and behavioral patterns aimed at survival 
in dynamic environments such as oceans or biological fluids ~\cite{pedley1992hydrodynamic,fenchel2002microbial,kiorboe2001marine}. 
Inspired by motile microorganisms, researchers have been developing artificial particles that feature similar swimming behavior either in a predefined way or by active control of their dynamical actions~\cite{lauga2009hydrodynamics,ebbens2010pursuit,ghosh2009controlled,mair2011highly,fischer2011magnetically,wang2012nano,gazzola2014reinforcement,gazzola2016learning}.  
One long-term goal is to engineer micro-robots that can swim toward specific target locations and autonomously perform complex tasks, with potential   biofluidic and pharmaceutical applications~\cite{nelson2010microrobots,gao2014environmental,abdelmohsen2014environmental,patra2013intelligent}.  
Another important goal suggested by behavioral studies is to understand  the origin of different navigation strategies in various fluid environments~\cite{reddy2016learning}. 
 
In this paper, we continue the study  presented in Ref.~\cite{colabrese2017flow}. The previous target was to find efficient motility strategies for a simple model of microswimmers in a two-dimensional flow via {\it reinforcement learning algorithms}. 
These algorithms are suitable to find optimal or approximately optimal  behaviours for performing a given pre-selected task.
The investigated simple model considers smart active microswimmers able to sense some basic cues from their environment. Furthermore, the particles have the ability to exert a control over some internal parameters that govern the preferred swimming direction with which they tend to align.
Using reinforcements from the environment the particle should find an efficient strategy to alter the internal parameter in order to achieve some long-term goal.
In Ref.~\cite{colabrese2017flow} we have shown that with a suitable choice of sensory inputs and actions, it is possible to find  very efficient strategies to escape from fluids traps, at least for two-dimensional flows with different degrees of complexity.
In this work, we start from the same setup of Ref.~\cite{colabrese2017flow} and we expand to the more complex case of  particles
immersed in a three-dimensional flow given by a superposition of two stationary Arnold-Beltrami-Childress (ABC) helical velocity fields. 
The ABC flows are particularly relevant for fluid dynamics, because they are exact stationary solutions of the Euler equations with non-trivial chaotic Lagrangian properties. As a result, in such flow configurations even a simple tracer which follows the streamlines is chaotically advected in the whole flow volume. Here we ask the question whether the swimmers  can learn to move efficiently through the chaotic flow towards a given direction (say upwards).

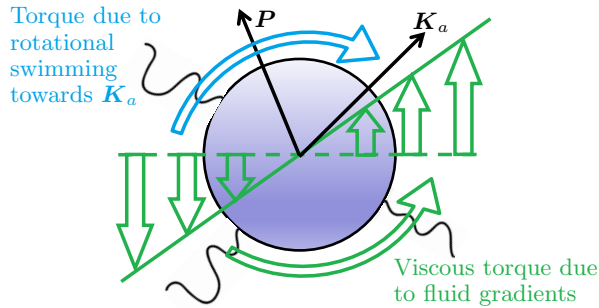
\begin{figure}[htbp]
	\centering
	\begin{overpic}[width=5cm]{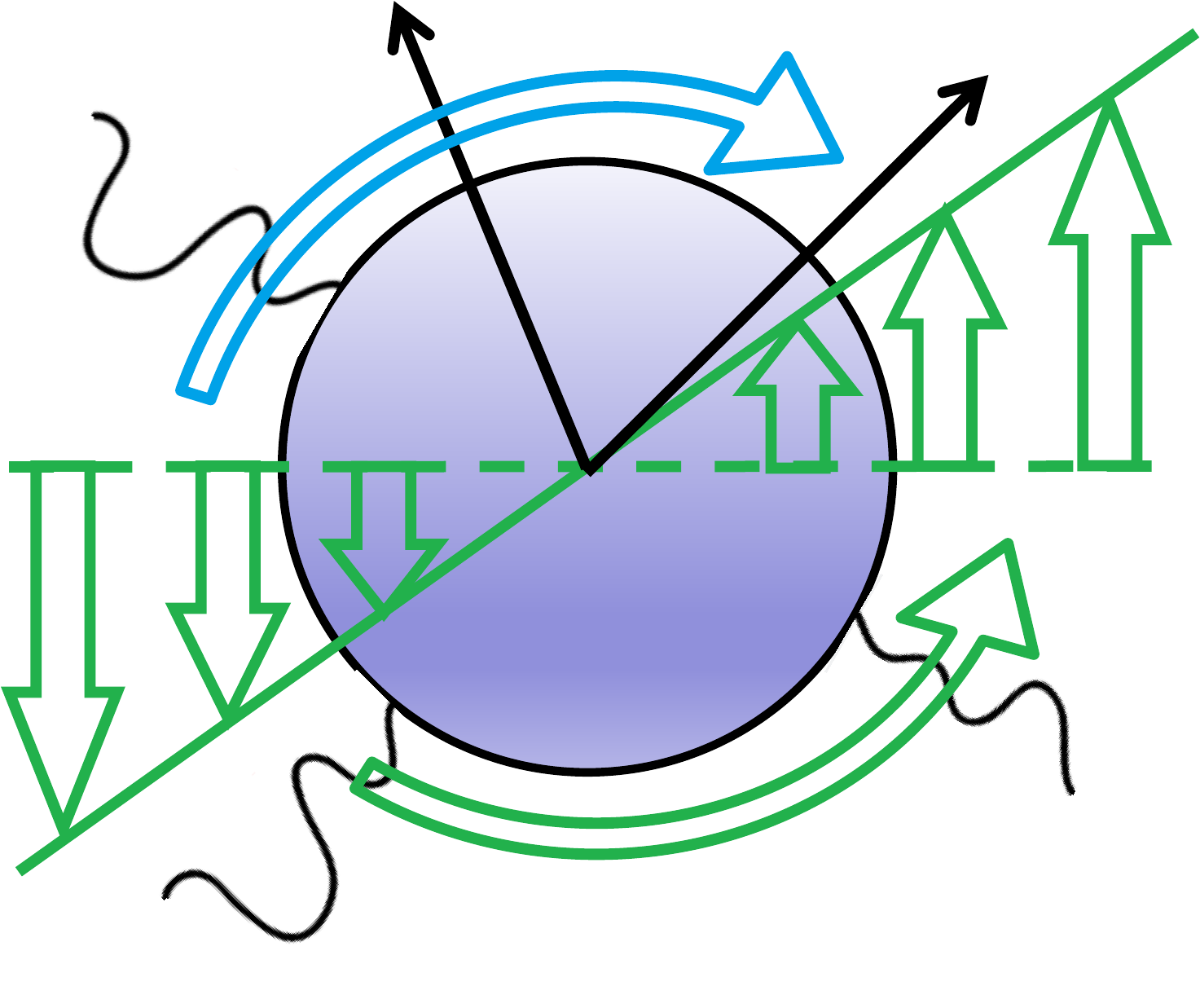}
		\put(37,79){$\ve{P}$}
		\put(79,78.5){$\ve{K}_a$}
		\put(-28,80){\color{MyBlue}Torque due to }
		\put(-28,73){\color{MyBlue}rotational}
		\put(-28,66){\color{MyBlue}swimming}
		\put(-28,59){\color{MyBlue}towards $\ve K_a$}
		\put(74,13){\color{MyGreen}Viscous torque due}
		\put(74,6){\color{MyGreen}to fluid gradients}
	\end{overpic}
	\caption{Sketch of the force balance on the body of the microswimmer. The smart microswimmer swims at all times with a constant swimming velocity with amplitude  $v_{\rm s}$ and with variable instantaneous direction $\ve{P}$.
		The orientation of $\ve P$ is determined by a competition between the viscous torque due to fluid gradients (green curved arrow), and the  torque controlled by the particle itself (blue curved arrow) such that it tries to rotate towards
		a direction $ \ve{K}_a $, with $a$ labelling one of the six possible orientations for the set-up  studied in this paper.
		The goal is  to learn the optimal policy  to swim upwards as fast as possible by dynamically selecting $\ve{K}_a$.}\label{fig:1}
\end{figure}

We consider an active particle that moves with constant speed in a direction that results from the competition between the shear-induced viscous torque and a torque that is controlled by the particle itself (see Fig. \ref{fig:1}).
In absence of flow,  the optimal strategy is to behave as a naive \textit{gyrotactic particle} which is constantly  reoriented by gravity to swim upwards.
In the presence of a flow, this choice may reveal quite inadequate. Depending on the particular realization and the properties of the swimmer, there is a competition between reorientations due to gravity and due to rotations from the vorticity of the flow.
This competition may hinder efficient ascent through the flow and even cause naive gyrotactic particles to remain trapped at a given depth~\cite{santamaria2014gyrotactic,durham2009disruption,durham2013turbulence}. 
Similar trapping effects are observed also in other contexts dominated by nonlinear coupling between particle and surrounding flow, for example for driven inertial particles in laminar flows \cite{sarracino2016nonlinear,cecconi2017anomalous}.
Using limited information about the regions of the flow they visit, the smart particles gradually develop a policy to escape from trapping regions and to find efficient pathways in order to globally move against gravity faster than naive gyrotactic swimmers.
Reinforcement learning provides a way to construct such efficient strategies  by accumulating good and bad experiences.
\section{Smart active swimmers}
\label{sec:swimmer}
\noindent
We consider neutrally buoyant, passive microswimmers that are small enough so that all inertial effects and feedback on the flow field $\ve u$ can be neglected~\cite{purcell1977life}.
The translational motion of an individual microswimmer is governed by flow advection  and by the swimming velocity with
constant magnitude $v_{\rm s}$ in the instantaneous direction $\ve P$: 
\begin{equation}\label{eq:Pos}
\dot{\ve{X}} = \textbf{u}  + v_{\rm s} \textbf{P} + \sqrt{2D_0}\ve\eta\,.
\end{equation}
Here $\dot{\ve X}$ denotes the time derivative of the position of the swimmer, $\ve X$.
In order to avoid structurally unstable dynamics, we have introduced a Gaussian white noise $\ve\eta(t)$ such that $\langle \eta_i(t)\eta_j(t') \rangle = \delta_{ij}\delta(t-t')$, where $D_0$ denotes the intensity of a small translational diffusivity.
The rotational motion of the particle is determined from the two torques illustrated in Fig.\ref{fig:1} and the corresponding change in the direction of $\ve P$ is given by~\cite{pedley1992hydrodynamic}:
\begin{equation}\label{eq:Dir}
\dot{\ve{P}} =  \dfrac{1}{2B}[\ve{K}_a- (\ve{K}_a\cdot \ve{P})\ve{P}] + \dfrac{1}{2}\ve\Omega \times \ve{P} +\sqrt{2D_{\rm R}} \ve\xi\,.
\end{equation}
Here $ \ve{K}_a$ is a given target direction, $ \ve{\Omega}\equiv\ve\nabla\times\ve u$ is the flow vorticity, $B$ is a parameter that determines the time scale of alignment between $\ve P$ and $\ve{K}_a$ in the absence of flow.
As we did for the translational motion, we have added also here a Gaussian white noise $\ve\xi(t)$ which introduces a rotational diffusivity via the small parameter $D_{\rm R}$ and it is such that the unit amplitude of $\ve P$ is preserved. 
In our case, the index $a$ of $\ve{K}_a$ takes a discrete set of values that corresponds to a finite number of possible target directions.
Naive gyrotactic particles have a single value $K_a=\hat z$ and always tend to align in the vertical  direction opposite to gravity.
Our smart active particle can choose its direction $K_a$ dynamically, depending on local cues measured along the trajectory,  as is explained in Section~\ref{sec:implementation}.

We denote the characteristic speed and magnitude of vorticity of the flow by $u_0$ and $\Omega_0$ respectively.
The dynamics in Eqs. (\ref{eq:Pos}) and (\ref{eq:Dir}) depends on two relevant dimensionless parameters.
First, the swimming number $\Phi\equiv v_{\rm s}/u_0$, that
quantifies the relative magnitude of the particle swimming speed to that of the flow.
Second, the stability number $\Psi\equiv B\omega_0$, that quantifies the relative magnitude of rotations due to the fluid vorticity compared to rotations due to the self-induced torque of the particle.
In the limit of $\Phi\gg 1$ and $\Psi\ll 1$ the flow can be neglected and the optimal strategy to navigate upwards will be simply given by the naive gyrotactic case, with ${\ve K}_a= \hat z$ at all times.
On the other hand, when  $\Phi$ is small, particles are mainly advected by the flow and navigation becomes hard.
Likewise, when  $\Psi$ is large, navigation is hard because particles are strongly  re-oriented  by the fluid vorticity.
In this work we will focus on microswimmers with parameter combinations in the regime of small $\Phi$ and large $\Psi$ where navigation is hard.
Typical values of $\Phi$ and $\Psi$ encountered in natural planktonic microswimmers in the ocean range from $ \Psi\sim 1$--$50$ and $\Phi\sim 0.03$--$1$ for surface water and $ \Psi \sim 0.01$--$50$ and $\Phi\sim 0.03$--$10$ for the very deep sea~\cite{gustavsson2016preferential}. 
\section{The three-dimensional chaotic flow}
\label{sec:flow}
\begin{figure}[htbp]
	\centering
	\includegraphics[width=5cm]{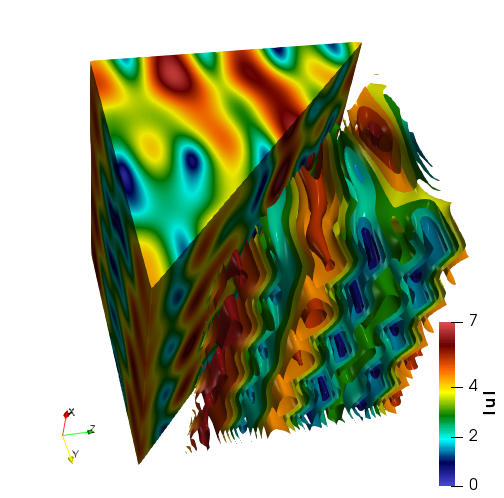}
	\caption{Colour-coded representation of the spatial dependence of the magnitude of the velocity, $|\ve{u}|$, for the superposition of two ABC flows considered in our study.}\label{fig:lsABC}
\end{figure}

\noindent
The  ABC family of flows is characterised by having vorticity perfectly aligned or anti-aligned with velocity at each point in
the three-dimensional space.
Even though the flow has a simple Eulerian representation, it has highly non-trivial Lagrangian properties.
In order to make navigation through the flow more complicated, we decided to explore a  less symmetric case by  taking
a linear superposition of two ABC flows:
%
\begin{eqnarray}
u_x&=& C\cos y + A\sin z + \widetilde{C} \cos 4(y + \Delta_1) + \widetilde{A} \sin 4(z + \Delta_2) \nonumber\\
u_y&=& A \cos z + B \sin x + \widetilde{A} \cos 4(z + \Delta_2) + \widetilde{B} \sin 4(x + \Delta_3)\nonumber\\
u_z&=& B \cos x + C \sin y + \widetilde{B} \cos 4(x + \Delta_3) + \widetilde{C}\sin 4 (y + \Delta_1), \nonumber
\end{eqnarray}
with parameters
\begin{eqnarray}
A = 1.5;\quad \widetilde{A} &=& 2.0;\quad \Delta_1 = 0.9; \nonumber\\
B = 0.5;\quad \widetilde{B} &=& 1.0;\quad \Delta_2  = 0.4 ; \nonumber\\
C = 2.5;\quad \widetilde{C} &=& 0.3;\quad \Delta_3 = 2.9.
\end{eqnarray}
For  what concerns the results shown in this paper, the exact choice of the linear superposition is not important. We have checked that with other similar values of the parameters all our results are qualitatively the same.
To visualize the spatial complexity of the flow we show the magnitude of the velocity field in Fig.~\ref{fig:lsABC}. 
\section{Implementation of reinforcement learning}
\label{sec:implementation}
\noindent
The reinforcement learning framework for an agent (the smart  microswimmer) has three basic components.
First, the {\it state} $s$ denotes what the agent senses.
The state $s$ is an element of $\mathcal{S}$, where $\mathcal{S}$ denotes a finite set of distinct states the agent is able to recognise.
Second, the agent can choose an {\it action} $a$.
The choice of action determines the dynamical evolution of the particle.
It is chosen depending on the current state and can be selected from a finite set of possible actions $\mathcal{A}$.
The final component is the {\it reward} $r$.
It is evaluated after the particle has interacted with the environment using an action $a$ determined by the state $s$.
It quantifies the immediate success or failure of the action chosen for the current state to reach the targeted  goal.

\begin{figure}[htbp]
	\centering
	\includegraphics[width=5cm]{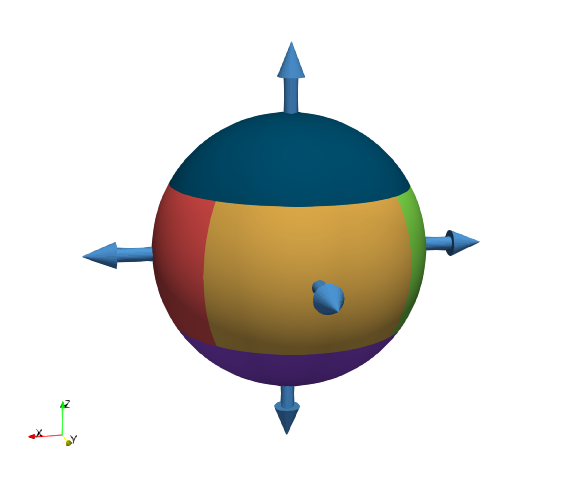}
	\caption{Sketch of the six angular sectors into which the instantaneous directions $ \ve{P}$ are mapped to define the set of states, ${\cal S}_p$.
		The unit sphere is divided into three segments separated by the planes $z=-0.5$ and $z=0.5$. The segment $-1\le z<-0.5$ corresponds to the direction $-\hat z$. The region $0.5< z\le 1$ corresponds to the direction $\hat z$. The remaining segment is divided into four equal-sized circular sections that are centered around the directions $-\hat x$, $\hat{x}$, $-\hat y$ and $-\hat y$.}
	\label{fig:sketch}
\end{figure}

In our implementation, particles can perceive only a coarse-grained representation of their current swimming direction $\ve{P}$ and of the local flow vorticity. We have chosen a state space $\mathcal{S}$ that is the product of two subsets, one identifying the coarse grained values of the $z$-component of the vorticity field, $\Omega_z$ and the second identifying a coarse grained value of the instantaneous swimming direction, $\ve P$. The discrete state space is $\mathcal{S} = \mathcal{S}_{\Omega_z} \times \mathcal{S}_{\ve{\mbox{\footnotesize{\boldmath$P$}}}} $, where $\mathcal{S}_{\Omega_z}$ are ten distinct equispaced intervals of $\Omega_z$ and where  $\mathcal{S}_{\ve{\mbox{\footnotesize{\boldmath$P$}}}} $ are the six angular sectors of the direction $\ve P$ depicted in Fig. \ref{fig:sketch}. As a result, we have in total $10\times6=60$ possible states.
The set of actions comprises six target swimming directions
${\ve K}_a \in \mathcal{A} = (\hat{x}, -\hat{x}, \hat{y}, -\hat{y}, \hat{z}, -\hat{z}) $.
Because we want the particles to go upwards, we let the reward be the net increase in the $z$-direction during any time interval the particle interacts with its environment.

The aim is to find an approximately optimal policy $\pi^*$  to choose an action $a$ given a certain instantaneous state $s$: $\pi^* : s \rightarrow a$~\cite{book:sutton}.
The algorithm constructs $\pi^*$ by the introduction of intermediate strategies $\pi_n$ during the $n$-th learning step, such that $ \lim_{n \rightarrow \infty} \pi_n = \pi^*$.
The instantaneous  policy $\pi_n$ depends in a deterministic way on the entries of a {\it quality matrix} $Q_{\pi_n}(a_n,s_n)$ according to a greedy rule: for any given state $s$ we pick the action $a$ that corresponds to the largest value of $Q$. This is motivated by the fact
that $Q$  quantifies the expected discounted sum of future rewards, conditioned on the current state $s_n$, action $a_n$ and policy $\pi_n$:
\begin{equation}
\label{eq:return}
Q_{\pi_n}(s_n,a_n) = \langle r_{n+1}\rangle + \gamma \langle r_{n+2}\rangle +
\gamma^2 \langle r_{n+3}\rangle + \cdots\,.
\end{equation}
Here the future rewards are given by the change in $z$ experienced between any two changes of state: $$r_{n+1}= z(s_{n+1})-z(s_n)$$
and $\gamma$ is the {\it discount factor} with $ 0 \leq \gamma < 1$. It quantifies the importance of future rewards.
For a value close to zero, only the next few rewards are taken into account, resulting in a short-sighted policy that greedily maximises the immediate reward, but may fail to reach a long-term goal.
For a value close to unity, we have a far-sighted optimization where later rewards give significant contributions, possibly allowing for long-term goals to be reached.
The one-step Q-learning algorithm that we use provides a way to update the matrix $Q$ such that for large $n$ we find an approximately optimal policy $\pi^*$ under suitable conditions~\cite{book:sutton}. The rule is that at any state change, $s_n \rightarrow s_{n+1}$,
we  update the value of $Q$ for the old state-action pair $(s_n,a_n)$  according to the reward, $r_{n+1}$,  and to the expected future rewards when being in the new state $s_{n+1}$ according to the current policy:
\begin{align}\label{eq:Q}
Q(s_n&, a_n) \leftarrow   \\
& Q(s_n, a_n) +   \alpha [r_{n+1} + \gamma \max_{a}Q(s_{n+1}, a) - Q(s_n, a_n)]\,. \nonumber
\end{align}
Here $0< \alpha <1$ is a free parameter which controls the learning rate~\cite{book:sutton}. It is possible to show~\cite{book:sutton} that for Markovian systems $ Q \rightarrow Q^{*} $ with probability $1$ if $ \alpha $ is taken to decay slowly to zero and the action selection rule samples each action an infinite number of times. 
The system considered here is not Markovian, but, as illustrated in Section \ref{sec:result}, the rule~(\ref{eq:Q}) still produces approximately optimal policies for large values of $n$.

Training sessions are organised  into subsequences, called episodes, $E$, with $E=1,\dots,N_E$ where $N_E$ is the total number of episodes
of each session. The first episode in each session starts with random position and orientation for the particle and with an optimistic initial strategy,  i.e. all entries of $Q$ are very large.  This is done in order to enhance  exploration in the space of the state-action pairs. At the end of each episode, we restart at another random position and orientation for the particle, but we keep the final $Q$ matrix from the previous episode. In this way, every time we restart a new episode we have some exploration due to the random restart
but we keep accumulating experience concerning the policy. In our particular case, an episode lasts until $T=8000$, where $T$ is a given
physical time.
\section{Results}
\label{sec:result}
\begin{figure}[htbp]
	\centering
	\includegraphics[scale=.6]{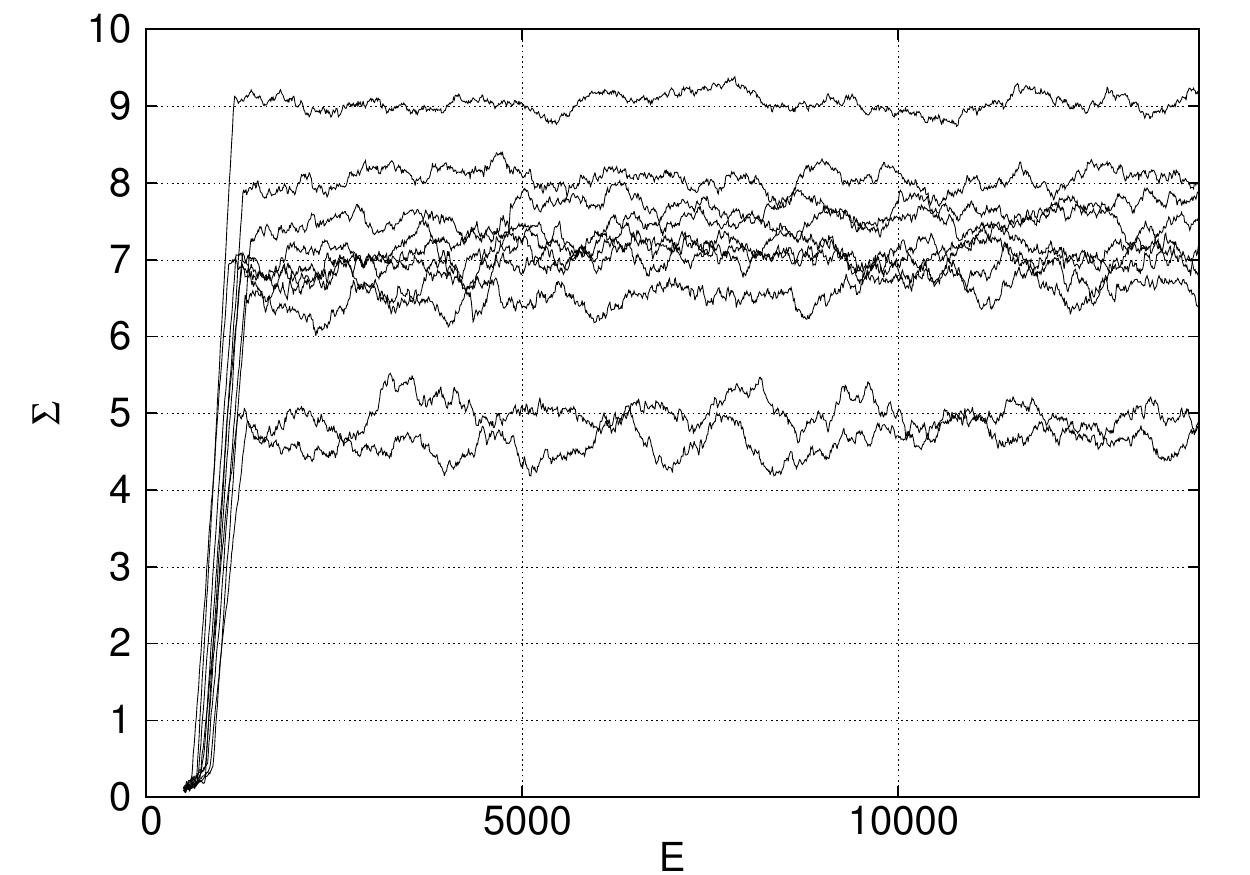}\caption{Dependence of the normalized total vertical distance $\Sigma(E)$ vs $E$ during 10 different independent training sessions (black curves). The value of $\Sigma(E)$ that is visualized is averaged  over  a window of 500 consecutive episodes to smooth out fluctuations. Notice that in all cases the smart particles are able to learn a strategy that outperform the naive case after $\sim  1000$ episodes. The small scatter among the optimal final strategies is due to the limited exploration introduced in the protocol (see the conclusions for a discussion about how it is possible to further improve these results using an $\epsilon$-greedy strategy). Parameters: $\Psi=9.5$, $\Phi=0.03$, $D_0\omega_0/u_0^2=0.004$, $D_{\rm R}/\omega_0=0.005$, $\alpha=0.1$, $\gamma=0.999$}
	\label{fig:gyroReturn}
\end{figure}

\begin{figure}[htbp]
	\centering
	\includegraphics[scale=0.51,trim={10mm 115mm 0mm 11mm}]{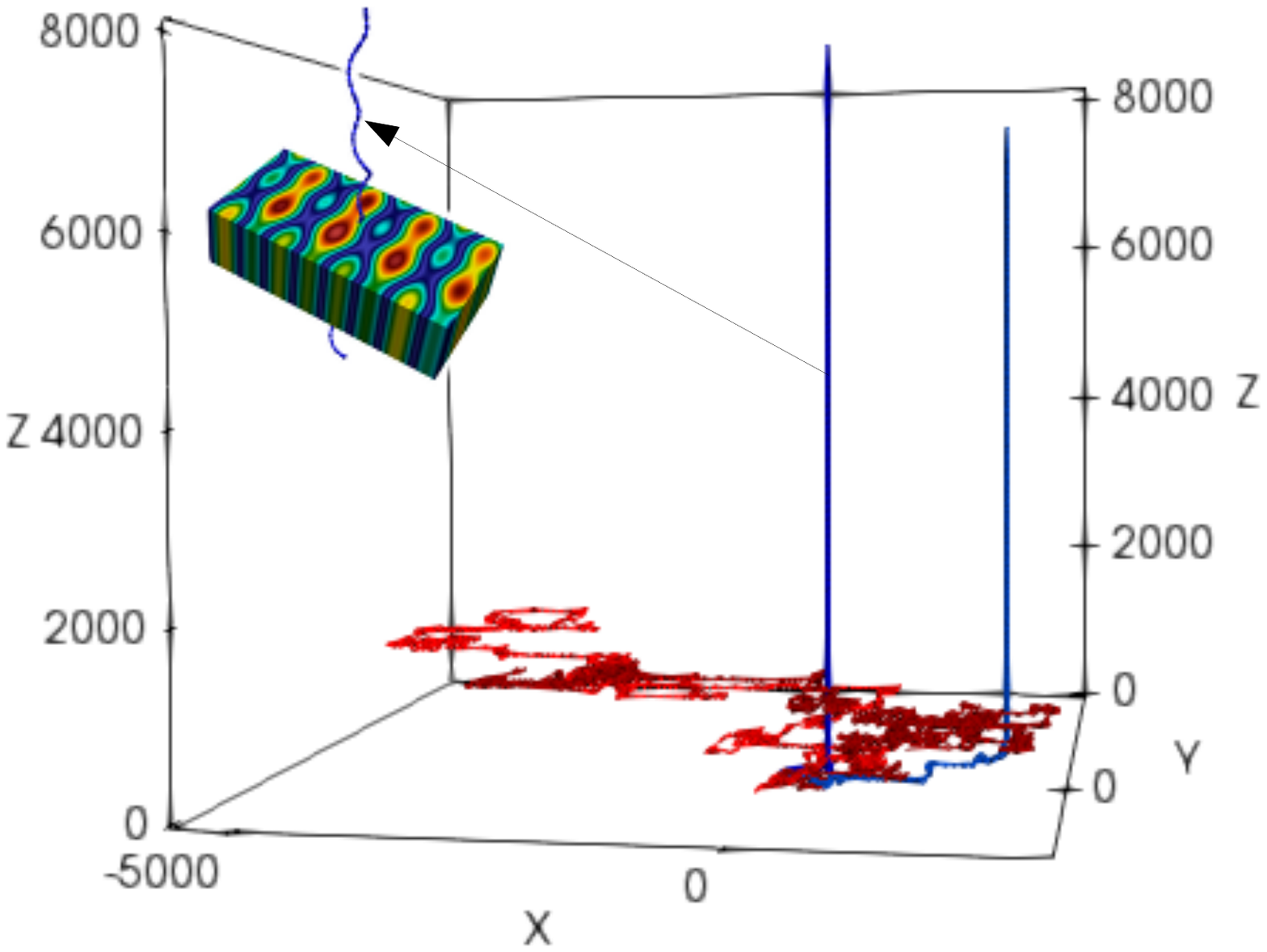}
	\caption{Three-dimensional evolution of two representative trajectories for smart active particles (blue) compared with two trajectories for naive gyrotactic particles (red). The inset shows a portion of a smart particle trajectory inside an elevator region. The $ \Omega_z $ field is colour-coded.} 
	\label{fig:learnProcess}
\end{figure}

\begin{figure}[htbp]
	\centering
	\includegraphics[scale=.6,trim={0mm 5mm 0mm 20mm}]{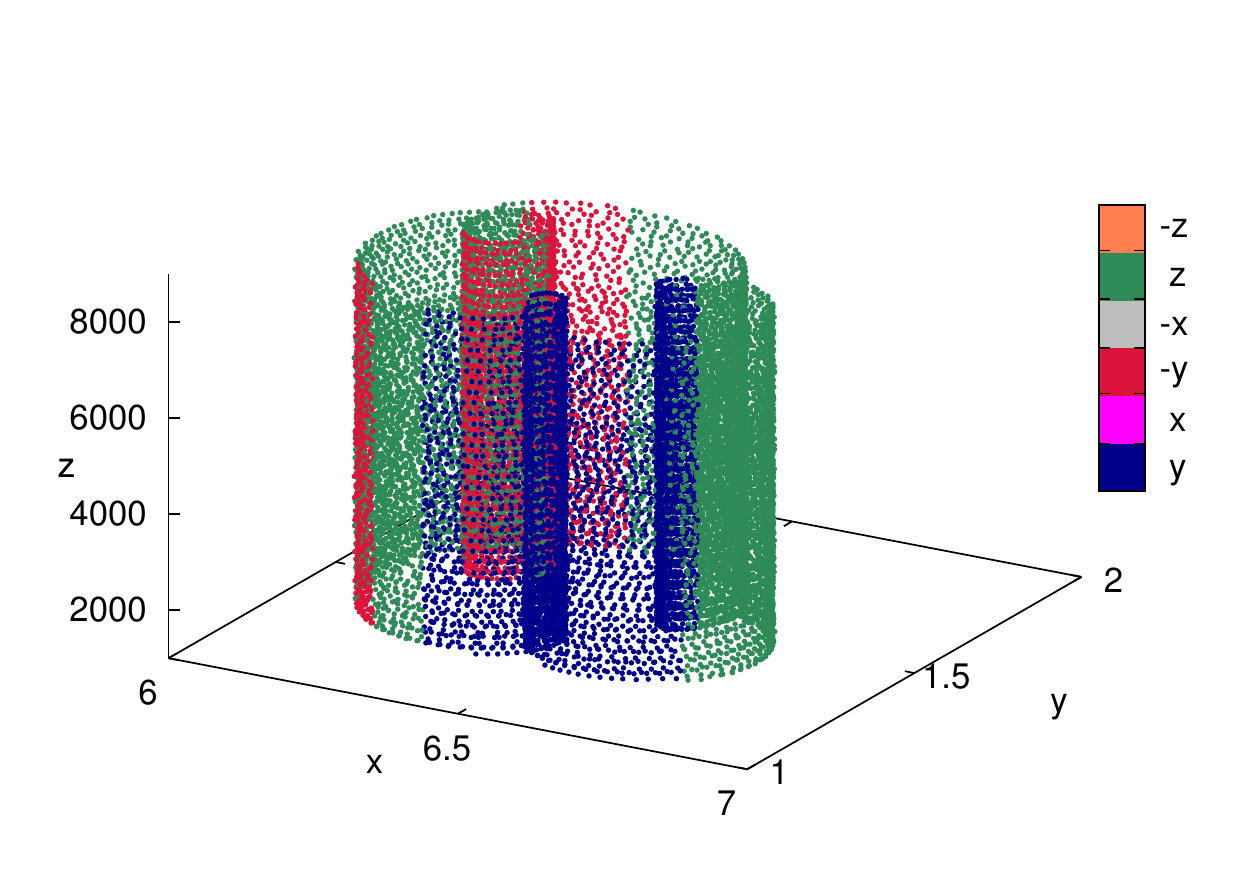}
	\caption{ Three-dimensional rendering showing how the six actions, $\ve{K}_a$, are taken in the elevator region  by smart particles following one of the approximately optimal policies obtained in Fig.~\ref{fig:gyroReturn}.}
	\label{fig:gyrotraj}
\end{figure}

\noindent
In order to quantify the ability of the smart particle to learn during a training session, we monitor the total vertical distance travelled by the swimmer during each episode $\Delta Z (E)= z(T) - z(0)$, normalized by the same quantity for the naive gyrotactic particle, the latter averaged over many different initial conditions:
\begin{equation}
\Sigma (E) = \frac{\Delta Z}{\langle \Delta Z_g \rangle}\,.
\end{equation}
Fig.~\ref{fig:gyroReturn} shows  the evolution  of $\Sigma$ vs $E$ during 10 independent training sessions. 
The blind policy at $E=0$ gives $\Sigma(E) \sim 0$ because the particles are as likely to go down as up before training. After a number of episodes, $E \sim 500$, the learning gain increases to eventually settle at an approximately optimal level. Which level is reached depends on the history of states, actions and rewards the particle encounters. These depend on the initial conditions and on the white noises in Eqs.(\ref{eq:Pos}) and (\ref{eq:Dir}). All training sessions in Fig.~\ref{fig:gyroReturn} are highly successful, outperforming the naive gyrotactic swimmer by obtaining an upward drift that is better by a factor between five and nine. In Fig.~\ref{fig:learnProcess} we show a visual comparison between representative trajectories: smart particles (blue trajectories) go upwards faster than naive particles (red trajectories) once they find the elevator regions in the flow. In order to quantify the learning gain, we perform also an \textit{exam} phase by using a fixed approximately optimal $Q$-matrix obtained at the end of one of the $10$ training experiments shown in Fig.~\ref{fig:gyroReturn}. In the exam, we evaluate the discounted return along one episode starting from the stable attractor of the dynamics, i.e. after that particles have forgotten their initial conditions:
\begin{equation}
R(E) = \sum_{n=1}^{N_s}r_n\gamma^n,
\end{equation}
where $ N_s $ is the total number of state changes for that episode and it is of the order of a few thousand.
%
In Table~\ref{tab:gyroReturn} we show the discounted returns averaged over $ 1000 $ trajectories  for smart $\langle R(E)\rangle $ and naive particles $\langle R(E)_g\rangle$, for three points in the parameter space. For the considered cases we found that smart particles show improvement over the naive gyrotactic ones, as pointed out in the third column of Tab.~\ref{tab:gyroReturn}, in which the ratio between the respective discounted returns is reported .
\begin{table}[htbp]
	\centering
	\begin{tabular}{|cccc|}
		\hline
		$(\Psi,\Phi)$ & $\langle R(E) \rangle$ & $\langle R(E)_g \rangle$ & $\langle R(E) \rangle/ \langle R(E)_g \rangle$\cr
		$(2.8,0.3)$ & 1026.4&359.5 & 3.0 \cr
		$(28.5,0.08)$ & 174.5& 85.4& 2.9\cr
		$(9.5,0.03)$ & 376.5& 26.4& 14.3\cr
		\hline
	\end{tabular}
	\caption{Average discounted return $\langle R(E) \rangle$ for some different parameter combinations $\Psi$ and $\Phi$ both for smart and naive particles. The last parameter combination corresponds to the data in Fig.~\ref{fig:gyroReturn}
	}\label{tab:gyroReturn}
\end{table}
In order to show the non-trivial policy that the smart particles are able to learn, we  visualize in Fig. \ref{fig:gyrotraj} the optimal actions taken in different spatial regions during the ascent through an {\it elevator}.
While the instantaneous direction of the smart particle always points upwards (it moves within the angular sector of the state $ \hat{z}$), the preferential direction changes depending on the $ \Omega_z $ states crossed by the particle.
\begin{figure}[htbp]
	\centering
	\includegraphics[scale=.75]{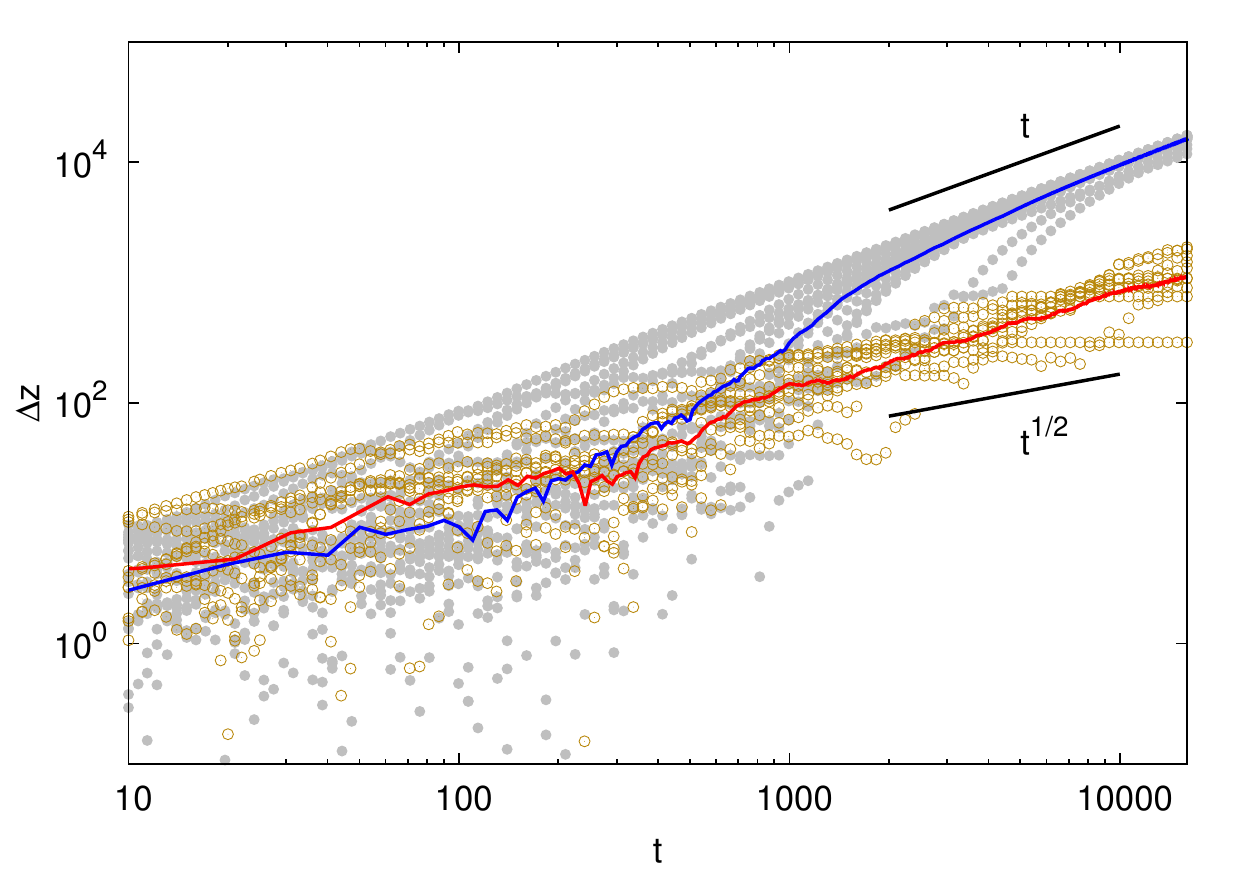}
	\caption{Logarithmic plot of the square root of the vertical mean square displacement, $\Delta z(t) = (\langle (z(t) -z(0))^2 \rangle)^{1/2}$, as a function of time for the  smart active particles (blue) and for naive particles (red).
		The average is over  $10$  trajectories also shown in the figure (grey filled circles for smart particles and yellow empty circles for naive particles). Notice that around $t \sim 1000$ the smart particles start to systematically outperform the naive case, indicating that this is the typical time needed to find the elevator regions inside the volume.
	}\label{fig:logscale}
\end{figure}
To quantify the long-term gain  in the upward drift, we show the square root of the mean square displacement:
\be
\Delta z(t) = (\langle (z(t) -z(0))^2 \rangle)^{1/2}
\ee
in a logarithmic scale in Fig.~\ref{fig:logscale}.
For large times, the smart particles have a quasi-ballistic behaviour with $\Delta z (t) \sim t$, while naive gyrotactic particles are drifting upwards much slower.
The data refers to the exam phase, in which we use a policy derived from a constant $Q$-matrix taken from the final episode of a training session.
It is interesting to notice that while $\Delta z$ for naive particles has a constant slope when plotted against time,
$\Delta z$ for smart particles change slope after the characteristic time needed to find an elevator region in the underlying flow.
It is not clear why this time scale is so large ($t\sim 1000$).
We expect that changing the learning framework, for example by choosing a reward that is beneficial for quick ascension, should result in a strategy that more quickly finds the elevator regions.

\section{Conclusions}
\noindent
In this paper we have extended the analysis done in Ref.~\cite{colabrese2017flow} to show how smart microswimmers
are able to learn approximately optimal swimming policies  to escape fluid traps and to efficiently swim upwards even in the presence of a complex three-dimensional underlying flow with chaotic trajectories.
We achieve this goal by applying a reinforcement learning algorithm, where the microswimmer adapts its strategy by learning from experience. In particular, we rely on the one-step Q-learning algorithm to converge to an approximately optimal policy in an iterative way. We benchmark the performance by comparing the ability to reach the goal (in our case to quickly move upwards) with the case of naive particles that cannot adapt their swimming direction. For all cases we tried, we found that the policies generated using reinforcement learning outperform the naive ones. The smart particles are able to find  elevator regions in the three-dimensional volume, allowing them to quickly ascend the flow, while naive particles are meandering around  with a slower mean vertical  velocity. In principle, even better strategies could be obtained by adopting an $\epsilon$-greedy algorithm, i.e. allowing for additional exploration during the learning phase by leaving to the particle a small probability $\epsilon$ to choose an action  different from the optimal one, at each time the state changes  ~\cite{book:sutton}. This is sometimes useful to further enhance exploration vs exploitation during the learning phase.

It is important to make clear that we are not interested at this stage to make a fully realistic model of either the particle dynamics, or the actual complexity of real fluid flows. 
At this stage we also avoid complications that may be present if one were to implement the reinforcement learning protocol online for physical microswimmers, such as time delays between sensing, actions and rewards.
Our goal is instead to propose a way  to develop novel strategies for designing smart devices. The reinforcement learning algorithm provides solutions to perform difficult tasks in complex environments.\\
Acknoledgements: SC and LB acknowledge funding from the European Research Council under the European Unions Seventh Framework Programme, ERC Grant Agreement No 339032.

\section{Authors contributions}
\noindent
All the authors were involved in the preparation of the manuscript.
All the authors have read and approved the final manuscript.


\end{document}